# Directional toroidal dipoles driven by oblique poloidal current flows in plasmonic meta-atoms

ARASH AHMADIVAND[*] AND BURAK GERISLIOGLU

*Department of Physics and Astronomy, Rice University, 6100 Main St, Houston, Texas 77005, United States*
*Corresponding author: aahmadiv@rice.edu*

**The toroidal dipole is an exquisite electromagnetic momentum apart from the classical multipoles family that can be localized and squeezed in an extremely tiny spot. This mode, in optical nanosystems, can be particularly distinguished as a non-radiating charge-current configuration, manifests in a head-to-tail fashion. Here going beyond conventional toroidal dipoles, we explained the possibility of the excitation of directional toroidal dipoles in plasmonic meta-atoms driven by controlling the oblique poloidal fluxes. We showed that the unique structural properties of the proposed asymmetric design lead to the formation of multiple toroidal modes with opposite poloidal current flows.**

Toroidization or toroidal polarization has been acknowledged as the description for condensed matter principle based on toroidal multipoles [1]. Technically, the residues from an electric octupole (**p**) and a magnetic quadrupole (**m**) can be combined to form an exquisite nonradiating closed-loop charge-current configuration, known as an axial toroidal dipole moment (**T**) [1,2]. Discovered, revealed and described for the first time by Zel'dovich [3], toroidal multipoles have been studied in the context of solid state physics [4], atomic and nuclear physics [5], and condensed matters [6]. In terms of the latter approach, time reversal ($t \rightarrow -t$) and space inversion ($r \rightarrow -r$) symmetries are the fundamentals of the toroidization principle for both vortex-like and polar configurations [2-4]. Such violations in the classical electromagnetic symmetries give rise to the formation of strongly squeezed vortex of a head-to-tail arrangement (toroidal dipole) with ultra-weak far-field radiation signature, also known as "ghost resonances" [4,7]. The major production of this interference is a nonzero contribution to the scattered far-field radiation pattern, can be described particularly by the summation of induced charge ($Q \propto \frac{d}{dr}(r \cdot j_l)$), transversal ($M \propto |r \times J|$) and radial ($T \propto |r \cdot J|$) currents as follows:

$$E_s = \frac{4\pi k^2}{c} \sum_{l,m} Q_{l,m}\psi_{l,m} + \sum_{l,m} M_{l,m}\phi_{l,m} + \sum_{l,m} T_{l,m}\psi_{l,m} \quad (1)$$

where $k$ is the wavenumber, $c$ is the speed of light in a vacuum, and $Q$, $M$, and $T$ correspond to the electric, magnetic, and toroidal multipoles components, respectively. Possessing properties similar to the dark-side of plasmons, toroidal dipole moment has received growing interest for several practical applications because of facilitating negligible radiative losses due to suppression of electric dipole [4,8,9]. On the other hand, the hybrid behavior, narrow lineshape, and ultrahigh sensitivity of toroidal dipole-resonant metamaterials to the environmental perturbations have led to emerging of several optical and biological-sensing devices with unique spectral properties [10-13]. Very recently, several innovative methods have been utilized to increase the tunability of toroidal meta-atoms and metamolecules to enhance the functionality of developed devices [14], such as integration of multipixel planar plasmonic toroidal meta-atoms and phase-changing materials [15] to design hybrid and fast modulators [16,17].

The excitation of a dynamic toroidal dipole accompanies with the formation of a closed-loop configuration on the surface of a torus as a "poloidal current". The direction of the poloidal current strongly relies on the magnetic field direction in the adjacent pixels of a given plasmonic unit cell, however, the direction of the poloidal currents cannot be effectively controlled and functionalized. In addition, poloidal currents are the significant and distinct manifest for the excitation of toroidal moments, therefore, having control on the direction, quality, and squeezing of this feature leads to having functional, and directional toroidal resonances. As a leading work, Bellan *et al.* [18] have analytically shown the variations in the vorticity model of flow can be driven by poloidal currents using magnetohydrodynamic models equations. This method explained the influence of axisymmetric poloidal flux surfaces on the axially directed vorticities in plasma systems. However, the influence of the poloidal fluxes and currents on the directivity of toroidal moments in metamaterials has not been investigated yet.

In this Letter, we study on the multiresonant toroidal plasmonic meta-atom to support directional toroidal resonances depending on the oppositely spinning (clockwise and counterclockwise rotation) oblique poloidal currents across the near-infrared region (NIR) of the spectrum. Using well-engineered, asymmetric multipixel aluminum (Al) nanoresonators in a single unit cell, we demonstrated the physical mechanism behind the excitation of directional toroidal dipoles through the formation of opposite poloidal currents in a given asymmetric plasmonic meta-atom. Our surface current density and optical binding force calculations strongly validate the directional toroidal moment claim. Moreover, we show that the exquisite design of the structure leads to θ∼22° tilting in the polarization angle of the transmitted beam from the resonators.

Figure 1a shows the schematic of the proposed meta-atom located on a quartz substrate with the permittivity of ε=2.1 (not to scale). To model the unit cell in the current analyses, we used the empirically measured and defined Al dielectric function derived using ellipsometric data by assuming the presence of a thin diametric (∼ 2 nm) alumina ($Al_2O_3$) layer as a coverage coating the structure [19]. This was achieved based on mixing the tabulated values of Al and $Al_2O_3$ layers using Bruggeman approach, reported by Palik [20]. The exact geometrical sizes are reported in Fig. 1b, obtained by judiciously selecting and simulating various parameters for each resonator and these values are assumed to be fixed along the work. To define the exact geometries, analyze the spectral response, and tune the poloidal current flow direction for the proposed asymmetric toroidal nanostructure, we utilized Finite-Difference Time-Domain (FDTD) solver (Lumerical 2018). In the 3D simulations, the spatial grid sizes were set to Δ=1 nm in all axes and the time step was set to $dt$∼0.019 fs to establish the Courant stability factor close to ∼0.99 [21-33]. The incident pulse was assumed to be a plane wave and the workplace was encompassed by periodic boundaries to resemble a metasurface structure with infinite number of unit cells in arrays along both $x$ and $y$-directions, while the normal beam propagation direction ($z$-axis) is bounded with 32 perfectly matched layers (PML).

The simulated normalized multipole decomposition intensity profile for the electromagnetic response of the proposed plasmonic nanostructure is plotted in Fig. 1c. The polarization of the incident beam is transverse ($\varphi$=90°). The spectra is obtained based on the radiation transmitted and reflected by the planar arrangements of infinite number of scatterers, defined by the approach provided by Marinov *et al.* [34] for toroidal meta-atoms. The plotted curves reveal the excitation of two distinguished resonance located at λ∼975 nm and ∼1450 nm, correlating with the toroidal dipole moments (**T**). According to the toroidization concept of the optical physics, the projected toroidal radiation field ($\vec{E}_s$) includes the signature of both electric and magnetic components as described below [35]:

$$\vec{E}_s = \left(\frac{\mu_0 c^2}{2A^2}\right)\exp(-ikr)$$
$$\left[-ik\vec{p}_P + ik\hat{r}\left(\vec{m}_P - k^2\vec{m}_P^{(1)}/10\right) - k^2\left(\vec{T}_P - k^2\vec{T}_P^{(1)}/10\right)\right]$$
**(2)**

where $\mu_0$ is the magnetic permeability of free space, $A$ is the area of the unit cell, $r$ is the perpendicular distance of the monitor from the metasurface, $\hat{r}$ is the unit vector perpendicular to the unit cells array. In addition, $\mathbf{m}^{(1)}$ and $\mathbf{T}^{(1)}$ denote the mean-square radii of magnetic and toroidal dipoles, respectively, which are considered for the unit cells with finite geometries. It should be underlined that we neglected the influence of the multipolar toroidal modes due to having super weak far-field signature, hence, here the dominant multipoles are shown merely. Obviously, at both resonant wavelengths (λ∼975 nm and 1450 nm), the toroidal modes become dominant due to suppression of all magnetic (**m**) dipole and electric (**p** and **Q**) multipoles. It should be noted that the noticeable amplitude of the quadrupolar electric and dipolar magnetic modes is because of normalization of the profile. Figure 1d validates the consistency between the calculated multipole scattering intensity and transmission/reflection spectra for the proposed metasurface. Accordingly, the optically induced toroidal dipoles act as leading moments consistent with the previous studies in Fig 1c. This plane also contains the transmission curve for the longitudinal polarization ($\varphi$=0°), which verifies the strong polarization-dependency of the unit cell. Figures 1e and 1f illustrates the electric-field (E-field) maps for the excitation of toroidal dipoles under transverse polarization excitation in a logarithmic scale. The noteworthy fact here is the confinement and intensification of the fields at the specific parts of the capacitive opening between the central and proximal resonators at different wavelengths. Clearly, for the toroidal dipole at 975 nm, the field concentration is in central part of the unit cell, and conversely, for the toroidal moment at 1450 nm, the fields are localized at the tips of the outermost of the meta-atom. This effect reveals the formation of opposite toroidal dipoles, and the reason will be explained in the following subsection.

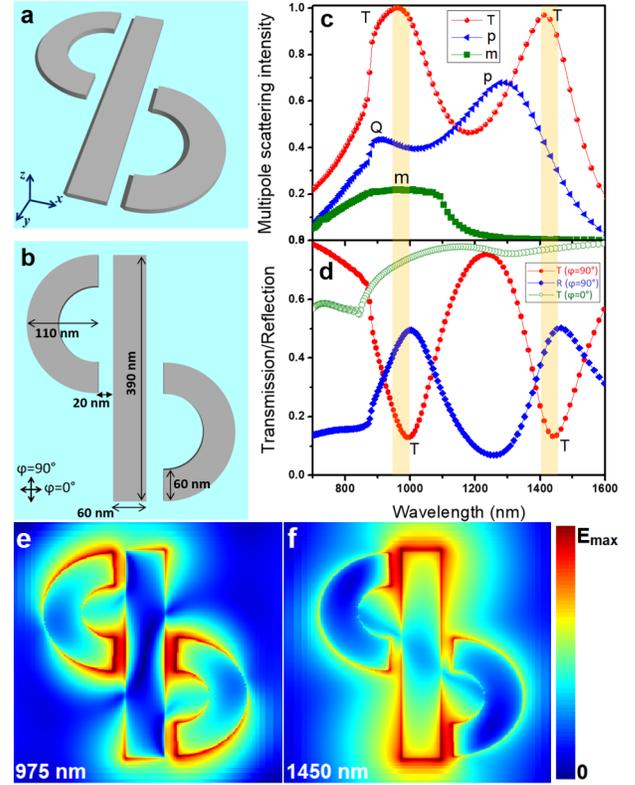

Fig. 1. a) Schematic for the design of the proposed Al meta-atom. b) The corresponding geometrical sizes for the proposed structure. The corresponding geometrical sizes are specified in the picture for all parameters. c) Normalized multipole scattering intensity from the scatterer for toroidal (**T**), electric (**p**) and magnetic (**m**) components. d) Normalized transmission and reflection patterns radiated from metamaterial. e, f) The E-field maps for the toroidal dipole resonances in the meta-atom in logarithmic scale.

As discussed earlier, the induced surface current and magnetic fields direction in the proximal resonators in a multipixel meta-atom have fundamental and substantial impacts on the formation and direction of toroidal resonances. Figure 2a exhibits an artistic illustration for the formation zone and rotation of the toroidal field, squeezed between the proximal and central resonators. This graphic also allows to understand the formation of the oblique closed-loop arrangement around the central block and rotation of poloidal current in tilted fashion. In Figs. 2b and 2c, we computed and plotted the vectorial surface current density maps for both toroidal resonances at different wavelengths to validate the excitation of directional toroidal resonances enforced *via* opposite oblique poloidal currents fluxes. These planes clearly show the required mismatch between the induced magnetic fields in each structure and also illustrate how the direction of the toroidal dipole inversely changes due to the variations in the direction of the poloidal currant flow. The direction of the incident normal plane-wave is shown in the inset of these planes. Here, the direction of the induced current for each dipolar mode. This trend also has a significant influence on the field localization as shown in Fig 1. Relatively, for the mode at 975 nm, the surface current flow direction in peripheral resonators converges to the center of the unit cell, consistent with Fig. 1e. For the other mode at 1450 nm, the surface current direction on the proximal half-rings is in the direction to the outermost part of unit cell in both sides, consistent with Fig. 1f. This can be easily understood by plotting the cross-sectional (*yz*-plane) magnetic-field (H-field) vectorial boards. We depicted the direction of the opposite poloidal currents at different wavelengths of the transmission dips Figs. 2d and 2e. For the toroidal dipole at 975 nm, we monitored a counterclockwise rotation of poloidal current flux and conversely, for the toroidal mode around 1450 nm, the poloidal current spins in clockwise path. This phenomenon allows for the formation of directional toroidal dipolar moments in a single unit cell.

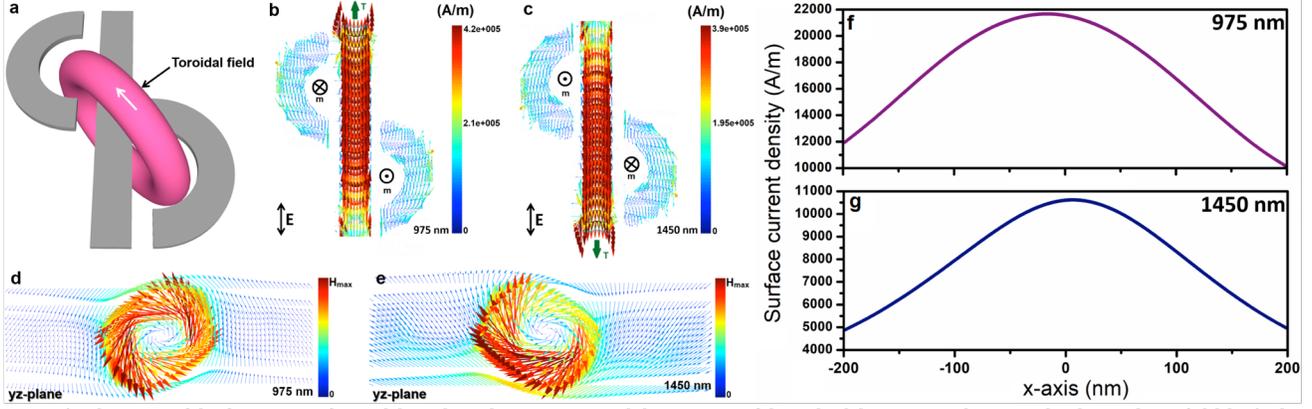

Fig. 2. a) Schematic of the formation of toroidal mode in the structure and the rotation of the poloidal current is shown in the donut shape field. b, c) The top-view vectorial surface current density maps for the directional toroidal modes. d, e) The cross-sectional *yz*-plane vectorial H-field maps for the excitation of toroidal field across the structure at two different wavelengths. f, g) The cross-sectional surface current density for both toroidal resonances at 975 nm and 1450 nm, respectively, in *yz*-plane.

Further analysis over the surface current density feature also enables us to compare the strength and quality of each induced dipolar mode. Figures 2f and 2g evaluate the surface current across the structure at the toroidal resonances wavelengths in *yz*-plane (cross-sectional view). Obviously, the induced current density at higher energies (975 nm) approximately two times stronger than the toroidal mode at low energies (1450 nm). Such a difference between the induced current density correlates with the strength of the magnetic moment in the toroidal mode position. On the other hand, we also computed the dephasing time for each induced resonances using the method developed based on the full width at half maximum (FWHM) computations as below [36,37]:

$$T = \frac{2\hbar}{FWHM} \quad (3)$$

where $\hbar$ is the reduced Planck's constant. Thus the dephasing time for the toroidal dipoles can be defined as approximately ~8.5 fs and 3.2 fs for the dipolar modes at 975 nm and 1450 nm, respectively. This shows longer lifetime of the toroidal mode at the shorter wavelengths due to having narrower resonance lineshape.

We also estimated the optical binding force (**F**) for the toroidal plasmonic scatterer to show the dominant enforcement arising from the poloidal current flux at the critical wavelengths (975 nm and 1450 nm) to determine the direction of the induced toroidal resonant moments. To this end, we employed the surface integration ($\mathbf{F} = \oint_S \langle \overleftrightarrow{\mathbf{T}} \rangle \cdot \hat{\mathbf{n}} ds$) using the time-averaged Maxwell stress tensor ($\overleftrightarrow{\mathbf{T}}$) [38]:

$$\langle \overleftrightarrow{\mathbf{T}} \rangle = \frac{1}{2}\text{Re}\left(\varepsilon\varepsilon_0 \mathbf{E}\mathbf{E}^* + \mu\mu_0 \mathbf{H}\mathbf{H}^* - \frac{\overleftrightarrow{\mathbf{I}}}{2}\left(\varepsilon\varepsilon_0 |\mathbf{E}|^2 + \mu\mu_0 |\mathbf{E}|^2\right)\right) \quad (4)$$

In the equations above, **n** is the normalized vector on the surface of a given meta-atom. Figure 3 demonstrates the computed binding force for a plasmonic unit cell, confirming the significant forces arising at the resonant wavelengths to push and define the direction of the toroidal moment.

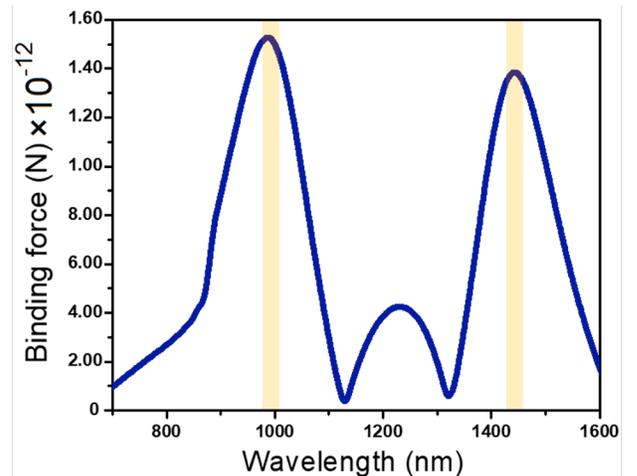

Fig. 3. The spectrum of the binding optical forces for the plasmonic toroidal meta-atom.

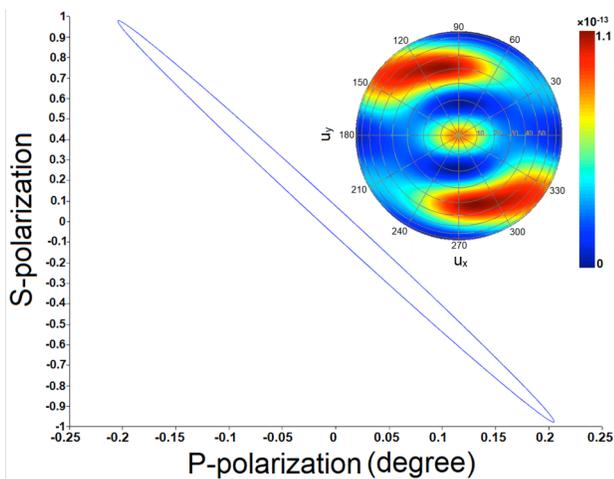

Fig. 4. The polarization ellipse for the oblique toroidal field angle. The inset is the far-field radiation pattern for the toroidal meta-atom.

Ultimately, we analyzed the specific feature of the asymmetric design and variations in the incident polarization angle *via* oblique signature of the toroidal field and its corresponding spinning symmetry. Figure 4 shows the polarization ellipse projected from the planar plasmonic toroidal metasurface. Here, the oblique angle of the polarization of the transmitted beam from the metasurface is calculated around $\theta \sim 22°$ (P-polarized axis) corresponds to the normalized incident beam (S-polarized axis). The inset is the far-field radiation pattern to validate the angle of the transmitted field polarization from the metasurface. This steering in the polarization angle of the transmitted electromagnetic beam can be utilized for light bending and steering applications, enhanced by toroidal modes.

In this work, we have numerically analyzed and explained the possibility of the excitation of polarization-dependent directional toroidal dipoles in plasmonic meta-atoms across the NIR of the spectrum. Our studies showed that optically driven oblique poloidal current flux have direct influence on the excitation of directional toroidal modes. Taking the advantage of the vectorial magnetic fields and surface current density profiles, we validated the formation of the directional toroidal fields. This claim was also verified by calculating the optical binding force at the resonant wavelengths. Finally, we calculated the polarization angle variations due to the oblique shape of the spinning toroidal fields across the multipixel unit cell. We envision that this work opens new avenues to the practical applications of the multiresonant toroidal meta-atoms for advanced nanophotonic applications from sensing to beam steering.

## References

1. A. E. Miroshnichenko, A. B. Evlyukhin, Y. F. Yu, R. M. Bakker, A. Chipouline, Arseniy I. K., B. Luk'yanchuk, B. N. Chichkov, and Y. S. Kivshar, "Nonradiating anapole modes in dielectric nanoparticles," Nat. Commun. **6**, 8069 (2015).
2. N. Papasimakis, V. A. Fedotov, V. Savinov, T. A. Raybould, and N. I. Zheludev. "Electromagnetic toroidal excitations in matter and free space," Nat. Mater. **15**, 263 (2016).
3. I. B. Zel'Dovich, "Electromagnetic interaction with parity violation," Sov. Phys. JETP **6**, 1184 (1958).
4. T. Kaelberer, V. A. Fedotov, N. Papasimakis, D. P. Tsai, and N. I. Zheludev. "Toroidal dipolar response in a metamaterial," Science 330, 1510 (2010).
5. Ceulemans, Arnout, L. F. Chibotaru, and P. W. Fowler, "Molecular anapole moments," Phys. Rev. Lett. **80**, 1861 (1998).
6. H. Schmid, "Some symmetry aspects of ferroics and single phase multiferroics," J. Phys. Cond. Mat. **20**, 434201 (2008).
7. W. Liu, J. Zhang, B. Lei, H. Hu, and A. E. Miroshnichenko, "Invisible nanowires with interfering electric and toroidal dipoles," Opt. Lett. 40, 2293 (2015).
8. M. Gupta, Y. K. Srivastava, and R. Singh, "A Toroidal Metamaterial Switch," Adv. Mater. **30**, 1704845 (2018).
9. A. Ahmadivand, B. Gerislioglu, and N. Pala, "Active control over the interplay between the dark and hidden sides of plasmonics using metallodielectric Au-Ge$_2$Sb$_2$Te$_5$ unit cells," J. Phys. Chem. C **121**, 19966 (2017).
10. T. Shibanuma, G. Grinblat, P. Albella, and S. A. Maier, "Efficient third harmonic generation from metal–dielectric hybrid nanoantennas," Nano Lett. **17**, 2647 (2017).
11. B. Gerislioglu, A. Ahmadivand, and N. Pala, "Tunable plasmonic toroidal terahertz metamodulator," Phys. Rev. B **97**, 161405 (2018).
12. A. Ahmadivand, B. Gerislioglu, P. Manickam, A. Kaushik, S. Bhansali, M. Nair, and N. Pala, "Rapid detection of infectious envelope proteins by magnetoplasmonic toroidal metasensors," ACS Sens. **2**, 1359 (2017).
13. A. Ahmadivand, B. Gerislioglu, A. Tomitaka, P. Manickam, A. Kaushik, S. Bhansali, M. Nair, and N. Pala, "Extreme sensitive metasensor for targeted biomarkers identification using colloidal nanoparticles-integrated plasmonic unit cells," Biomed. Opt. Express **9**, 373 (2018).
14. K. D. Heylman, D. Kevin and H. R. Goldsmith, "Photothermal mapping and free-space laser tuning of toroidal optical microcavities," Appl. Phys. Lett. **103**, 211116 (2013).
15. B. Gerislioglu, A. Ahmadivand, M. Karabiyik, R. Sinha, and N. Pala, "VO2-based reconfigurable antenna platform with addressable microheater matrix", Adv. Electron. Mater. 3, 1700170 (2017).
16. A. Ahmadivand, B. Gerislioglu, & N. Pala, "Active control over the interplay between the dark and hidden sides of plasmonics using metallodielectric Au–Ge$_2$Sb$_2$Te$_5$ unit cells," J. Phys. Chem. C **121**, 19966 (2017).
17. M. Gupta, Y. K. Srivastava, and R. Singh, "A Toroidal Metamaterial Switch," Adv. Mater. **30**, 1704845 (2018).
18. P. M. Ballen, "Vorticity model of flow driven by purely poloidal currents," Phys. Rev. Lett. **69**, 3515 (1992).
19. M. W. Knight, N. S. King, L. Liu, H. O. Everitt, P. Nordlander, and N. J. Halas, "Aluminum for plasmonics," ACS Nano **8**, 834 (2013).
20. E. D. Palik, in *Handbook of Optical Constants of Solids* 1998 (Academic Press, San Diego, CA).
21. B. Gerislioglu, A. Ahmadivand, and N. Pala, "Functional quadrumer clusters for switching between Fano and charge transfer plasmons", IEEE Photon. Technol. Lett. **29**, 2226 (2017).
22. B. Gerislioglu, A. Ahmadivand, and N. Pala, "Single- and multimode beam propagation through an optothermally controllable Fano clusters-mediated waveguide", IEEE J. Lightw. Technol. **35**, 4961 (2017).
23. B. Gerislioglu, A. Ahmadivand, and N. Pala, "Hybridized plasmons in graphene nanorings for extreme nonlinear optics", Opt. Mater. **73**, 729 (2017).
24. A. Ahmadivand, B. Gerislioglu, and N. Pala, "Azimuthally and radially excited charge transfer plasmon and Fano lineshapes in conductive sublayer-mediated nanoassemblies", J. Opt. Am. A **34**, 2052 (2017).
25. A. Ahmadivand, R. Sinha, M. Karabiyik, P. K. Vabbina, B. Gerislioglu, S. Kaya, and N. Pala, "Tunable THz wave absorption by graphene-assisted plasmonic metasurfaces based on metallic split ring resonators", J. Nanopart. Res. **19**, 3 (2017).
26. A. Ahmadivand, M. Karabiyik, R. Sinha, B. Gerislioglu, and N. Pala, "Tunable terahertz response of plasmonic vee-shaped assemblies with a graphene monolayer", Proceedings of the IEEE Progress in Electromagnetic Research Symposium (PIERS) 2387, Shanghai, China (2016), DOI:10.1109/PIERS.20167734979.
27. B. Gerislioglu, A. Ahmadivand, and N. Pala, "Optothermally controlled charge transfer plasmons in Au-Ge2Sb2Te5 core-shell dimers", Plasmonics (2018), DOI:10.1007/s11468-018-0706-6.

28. A. Ahmadivand, B. Gerislioglu, R. Sinha, M. Karabiyik, and N. Pala, "Optical switching using transition from dipolar to charge transfer plasmon modes in Ge2Sb2Te5 bridged metallodielectric dimers", Sci. Rep. **7**, 42807 (2017).
29. A. Ahmadivand, B. Gerislioglu, R. Sinha, P. K. Vabbina, M. Karabiyik, and N. Pala, "Excitation of terahertz charge transfer plasmons in metallic fractal structures", J. Infrared Milli. Terahz Waves **38**, 992 (2017).
30. A. Ahmadivand, B. Gerislioglu, and N. Pala, "Graphene optical switch based on charge transfer plasmons", Phys. Status Solidi RRL **11**, 1700285 (2017).
31. A. Ahmadivand, B. Gerislioglu, and N. Pala, "Large-modulation-depth polarization-sensitive plasmonic toroidal terahertz metamaterial", IEEE Photon. Technol. Lett. **29**, 1860 (2017).
32. B. Gerislioglu, A. Ahmadivand, and N. Pala, "Optothermally tuned charge transfer plasmons in Au-Ge2Sb2Te5 core-shell assemblies", MRS Advances **3**, 1919 (2018). DOI: 10.1557/adv.2018.258.
33. A. Ahmadivand, B. Gerislioglu, and N. Pala, "Optothermally controllable multiple high-order harmonics generation by Ge2Sb2Te5-mediated Fano clusters", Opt. Mater. **84**, 301-306 (2018). DOI:10.1016/j.optmat.2018.07.026
34. K. Marinov, A. D. Boardman, V. A. Fedotov, and N. Zheludev, "Toroidal metamaterial," New J. Phys. **9**, 324 (2007).
35. V. Savinov, V. A. Fedotov, and N. I Zheludev, "Toroidal dipolar excitation and macroscopic electromagnetic properties of metamaterials," Phys. Rev. B **89**, 205112 (2014).
36. T. Klar, M. Perner, S. Grosse, G. von Plessen, W. Spirkl, and J. Feldmann, "Surface-plasmon resonances in single metallic nanoparticles," Phys. Rev. Lett. **80**, 4249 (1998).
37. A. Ahmadivand, R. Sinha, B. Gerislioglu, M. Karabiyik, N. Pala, and M. Shur, "Transition from capacitive coupling to direct charge transfer in asymmetric terahertz plasmonic assemblies," Opt. Lett. **41**, 5333 (2016).
38. Q. Zhang, J. J. Xiao, X. M. Zhang, Y. Yao, and H. Liu, "Reversal of optical binding force by Fano resonance in plasmonic nanorod heterodimer," Opt. Express **21**, 6601 (2013).